# Mean-field theory of strongly disordered superconductors

Igor F. Herbut

Department of Physics, Dalhousie University, Halifax, Nova Scotia, Canada B3H 3J5

and

*Department of Physics, Simon Fraser University, Burnaby, British Columbia, Canada V5A 1S6

**Abstract:** Qualitative features of the mean-field theory of superconductivity in a strongly disordered system of fermions with short-range attraction are discussed. In this limit the effective theory at low energies is entirely bosonic, and I consider both the artificial infinite-range limit and the more realistic case of "nearest-neighbor" hopping of bosons between the localized states. In the infinite-range case the mean-field theory is exact, and the superconducting gap is uniform in space. There is a smooth BCS-BEC crossover with decrease in density, at weak enough disorder; at moderate densities, or larger disorder, the mean-field ground state is the BCS-like localized superconductor. In the latter case, the gap is highly non-uniform in space, but surprisingly is finite everywhere at $T < T_{MF}$. I find that the mean-field transition temperature $T_{MF} > 0$ always, and argue that the superconductor-insulator transition at $T = 0$ in models with net attraction between fermions is in the universality class of "dirty-bosons".

# 1 Introduction

The effect of disorder on superconductivity has been one of the fundamental issues in the field, which despite a lot of effort has still not been completely understood. It is well known that weak non-magnetic impurities do not affect much the transition temperature of a s-wave superconductor [1], [2], since one only needs to replace the plane-waves in the BCS variational ground state with the pairs of time-reversed exact eigenstates of the random single-particle part of Hamiltonian, to essentially preserve the result of the BCS theory. The eventual breakdown of the Anderson theorem with increase of disorder was studied in mid-eighties by several groups [3], [4], [5]. It was found that the BCS variational ground state continues to have lower energy than the non-superconducting state until the extreme limit of site-localization is reached, when the system becomes a gapless insulator of localized, tightly bound, singlet electron pairs. The inclusion of quantum fluctuations should lead to loss of long-range order in the ground state at a finite value of disorder, and the low-energy sector of the theory was argued to be equivalent to the system of disordered bosons with hard-core repulsion [4]. The question of critical behavior at the quantum ($T = 0$) superfluid-Bose glass transition in the system of dirty bosons has been of great interest [6], since it also represents a paradigmatic case of poorly understood interplay of localization and interactions. Recently, a systematic theory of the superfluid-Bose glass quantum critical point was developed based on the expansion around the lower critical dimension $d = 1$ for the transition [7].

Despite the success [8] of the bosonic theory of the quantum superconductor-insulator transition in naturally accounting for the observed scaling of resistivity and for the apparent universality of the critical conductivity, there is no consensus yet on the basic issue of its physical correctness. The bosonic theory assumes that electron pairs do not dissociate at the point of transition, but that it is only the long-range order in the phase of the superconducting order parameter that disappears. This is exactly what happens, in the strict renormalization group sense, at a finite temperature superconducting transition. A complementary view [9], supported by a different set of experiments [10], is that disorder enhances the effect of Coulomb interaction between electrons, to the point at which the net interaction becomes repulsive and the pairs break apart. Which of the two pictures should apply to a given experimental situation is at the present time far from being clear.

The purpose of the present paper is to clarify some of the conceptual issues behind the dirty-boson theory of the superconductor-insulator transition. In particular, I study the mean-field transition temperature in a model of strongly disordered fermions with a fixed net short-range attraction. Such a simplified model should presumably always show a quantum phase transition from a superfluid into a Bose glass of electron pairs, at $T = 0$ and with increase of disorder. One expects that the mean-field transition temperature, which should be understood as a crossover scale where the amplitude of the superfluid order-parameter starts forming, would remain finite at the critical value of disorder, i. e. that the superconducting gap does not collapse at the same point where the phase-coherence disappears (see Figure 1). This point is studied here by revisiting the mean-field theory of a disordered superconductor [4], but this time considering the problem at finite temperatures. Can the superconducting mean-field transition temperature $T_{MF}$ in a disordered Fermi system with purely attractive interaction between fermions vanish? Naively, one would allow this possibility at low densities of electrons, where the hard-core repulsion between bosonic electron pairs becomes less important, and the system should behave similarly to the gas of nearly independent bosons moving in a random external potential. The $T_{MF}$ would then correspond to the temperature where the Bose-Einstein condensation (BEC) into the lowest single-particle state takes place [11]. In this dilute limit, if the density of random eigenstates at the chemical potential at $T_{MF}$ (i. e. at the bottom of the random band) is finite, the $T_{MF}$ corresponding to the BEC would be zero [12], due to the disordering effect of a large number of zero-energy (Goldstone-like) modes. Essentially, $T_{MF}$ in this situation could vanish for the same reason as in the non-interacting gas of bosons in two dimensions. Motivated by this scenario I study first an admittedly artificial, but an instructive case of disordered lattice bosons with the infinite-range hopping, for which the mean-field theory is exact. In the infinite-range model the gap is uniform in space, and with decrease of particle density the transition temperature and the chemical potential indeed exhibit a clear BCS-BEC crossover if disorder is weak enough. Interestingly, $T_{MF}$ nevertheless always remains finite, and the answer to the question posed above is negative. In the limit of low-density and week disorder the chemical potential drops *below* the band of random energies, and right at $T_{MF}$ approaches the energy of the single, extended "bound" state, which then begins to become macroscopically occupied. The gap between this "bound" state and the rest of the band is finite at any disorder only for a uniform distribution of random energies; it collapses at strong disorder for any distribution that vanishes at the bottom, in which case the crossover to BEC of dilute bosons ceases to exist. Instead, at strong disorder one has the "localized superconductor" ground state [4], where the superfluid state is phase-coherently formed out of macroscopic number of localized states.

Retaining only the nearest-neighbor hopping in the effective bosonic theory leads to a more realistic and a more complicated problem, where the superconducting gap becomes strongly non-uniform in space. The non-linear mean-field equations that determine the values of the gap at different points in space can then be solved only numerically, however, some general statements on the nature of the superconducting solution can be made without explicitly finding the solution. Most importantly, the solution for the gap is either zero everywhere or nowhere in space. I discuss the mechanism behind the formation of the superconducting solution below $T_{MF}$, and the special role played by the sites with energies close to the chemical potential, which act as "seeds" of the long-range order [4]. In the limit of strong disorder these sites become sparse, and the gap exponentially quickly diminishes away from them, so that the system starts to resemble a granular superconductor with weak links connecting different granules. The transition temperature $T_{MF}$ is determined by the seed which is in the best local environment to first trigger the long-range order. Within the mean-field picture $T_{MF}$ is again always finite, and at $T = 0$ amplitude of the superconducting gap is everywhere positive, although with diminishing average value as disorder is increased. I also comment on the additional separation of energy scales which occurs at strong disorder, where the superconducting gap and the minimum single-fermion energy cease to be the same.

The paper is organized as follows. In the next section the effective low-energy bosonic Hamiltonian is derived starting from the disordered "negative-U" Hubbard model for electrons in the strong disorder limit. In Section 3 the BCS-BEC crossover at weak disorder and the "localized superconductor" at strong disorder in the model with infinite range hopping are discussed. The $T_{MF}$ in the case of nearest-neighbor hopping and strong disorder is calculated in the Section 4. Summary and connections to other works is given in the concluding section.

## 2 The low-energy bosonic theory

We will be interested in the system of lattice fermions in a random chemical potential, interacting via on-site attraction, defined with the standard negative-U Hubbard model:

$$H = -t \sum_{\langle i,j \rangle \sigma=+,-} c^\dagger_{i,\sigma} c_{j,\sigma} + \sum_{i,\sigma=+,-} V_i n_{i,\sigma} - U \sum_i n_{i,+} n_{i,-}. \quad (1)$$

Here $V_i$ is a random variable uniformly distributed in the interval $[-W/4, W/4]$, and $U > 0$. When $W/t$ is small, the eigenstates of the single-particle part of the Hamiltonian, given by the first two terms in Eq. 1, are close to Bloch waves, and the system has the usual BCS instability at small attraction $U$. In what follows I consider the opposite limit of a very strong disorder, when $W/t >> 1$. In the eigenbasis of its single-particle part the Hamiltonian (1) for a fixed realization of the random potential becomes:

$$H = \sum_{\alpha,\sigma} e_\alpha n_{\alpha,\sigma} - \sum_\alpha <\alpha,\alpha|\hat{U}|\alpha,\alpha> n_{\alpha,+} n_{\alpha,-} - \sum_{\alpha,\beta,\gamma,\delta,} <\alpha,\gamma|\hat{U}|\beta,\delta> c^\dagger_{\alpha,+} c_{\beta,+} c^\dagger_{\gamma,-} c_{\delta,-}. \quad (2)$$

In the strong disorder limit the exact eigenstates of the random single-particle part of the Hamiltonian approach the site-localized wave-functions, $|\alpha> \to |i>$, so $<\alpha,\alpha|\hat{U}|\alpha,\alpha> \approx U$, and the last term in the Eq. (2) which involves the overlap between different strongly localized states may be treated as a small perturbation. In the extreme site-localized limit when $t/W = 0$ the last term in Eq. (2) vanishes, and the ground state is simply

$$|\psi_0> = \prod_{i,(2V_i-U<0)} c^\dagger_{i,-} c^\dagger_{i,+} |0>. \quad (3)$$

Crucial observation is that the low-energy excitations above this ground state are purely *bosonic*: to create a pair of electrons at the next available energy level does not cost any energy, while to create a single electron there costs a finite amount of $U/2$ [4]. Despite the single-particle gap, the ground state does not have any off-diagonal long-range order characteristic of a superfluid, since all the states are either fully occupied with a pair of electrons or empty, and it, in fact, represents a localized insulator.

At strong disorder the last term in the Hamiltonian (2) is small, but finite. There are three different combinations of indices in this term that then give a dominant contribution: 1) the Hartree term, $\alpha = \beta \neq \gamma = \delta$, 2) the pair-breaking term, $\alpha = \delta \neq \beta = \gamma$, and 3) the pair "hopping" term, $\alpha = \gamma \neq \beta = \delta$.

For a zero-range interaction all three matrix elements are equal, and in the limit of strong disorder exponentially small compared to $U$. In contrast to the Hartree and the hopping terms 1) and 3), the term 2) annihilates any state that consists only of pairs of electrons, and therefore may be dropped in the low-energy sector of the theory, at energies much lower than $U$. One may then introduce the composite *bosonic* operators

$$b_\alpha = c_{\alpha,+}c_{\alpha,-}, b_\alpha^\dagger = c_{\alpha,-}^\dagger c_{\alpha,+}^\dagger \qquad (4)$$

and restrict the available Hilbert space by introducing the hard-core constraint as $(b_\alpha^\dagger)^2 = 0$, to account for the Pauli principle. The effective low-energy Hamiltonian in the strong-disorder limit is therefore:

$$H = \sum_\alpha E_\alpha b_\alpha^\dagger b_\alpha - \sum_{\alpha \neq \beta} J_{\alpha,\beta}(b_\alpha^\dagger b_\beta + 2 b_\alpha^\dagger b_\alpha b_\beta^\dagger b_\beta), \qquad (5)$$

with $E_\alpha = 2e_\alpha - U$, and $J_{\alpha,\beta} = <\alpha, \alpha|\hat{U}|\beta, \beta> = <\alpha, \beta|\hat{U}|\alpha, \beta>$. In the strong-disorder limit we can further simplify the effective Hamiltonian (5) as follows. First, the random lattice formed out of maxima of sharply peaked localized states $|\alpha>$ is not very different from the regular lattice we started with, and we will assume that they are identical. The difference can be absorbed into the overlap integrals $J_{\alpha,\beta}$. Second, the random matrix elements $J_{\alpha,\beta}$ will be assumed to be all the same, and equal to $J$ if the "sites" $\alpha$ and $\beta$ are "nearest neighbors", and to zero if they are not. As will be shown later, this does not affect qualitatively the main features of the mean-field theory. The only random variables left in the effective theory then are the site-energies $E_\alpha$, and the effective Hamiltonian describes the system of hard-core bosons hoping on a regular lattice in presence of random chemical potential, i. e. the system of "dirty bosons".

The reader may have noticed that the transformation of the Hamiltonian from Eq. 1 into Eq. 2 is completely general, and in principle may be performed even in the limit of weak disorder. In that case however, all the interaction terms in the Eq. 2 are of the same order, since the exact eigenstates are extended (or weakly localized in two dimensions) and the overlaps are numerous and large. So at weak disorder there is no clear separation of the Hamiltonian into the main part and the small perturbation, and the utility of the exact eigenstate representation is somewhat diminished. Proceeding nevertheless with the reduced bosonic Hamiltonian and assuming equal overlaps between the each pair of exact eigenstates leads to the Anderson theorem [1].

## 3  Infinite-range hopping

It is often useful to consider first the modification of the theory for which the mean-field approximation should be exact. For that purpose, assume that the hopping integral $J_{\alpha,\beta}$ in (5) is the same for *all* pairs $(\alpha, \beta)$ on the lattice. As discussed above, this assumption would be justified if disorder in the initial Hamiltonian (1) were weak and the system is three dimensional, so that all the eigenstates $|\alpha>$ were extended in space: the overlap integral $J_{\alpha,\beta}$ would then indeed be nearly independent of the states $\alpha$ and $\beta$. In the strongly disordered limit under consideration this is just a mathematical device. The effective low-energy Hamiltonian becomes

$$H = \sum_\alpha E_\alpha b_\alpha^\dagger b_\alpha - \frac{J}{M} \sum_{\alpha,\beta} b_\alpha^\dagger b_\beta, \tag{6}$$

where the second sum now is performed over *all* states $\alpha$ and $\beta$, $M$ is the total number of states (or lattice sites), and I dropped the Hartree term, which in the infinite-range model contributes only to an overall shift of the chemical potential. The hopping term is now of the form $O^\dagger O$, with the operator $O = \sum_\alpha b_\alpha$. Since $\langle O \rangle$ is an extensive quantity of order $M$, the standard mean-field decoupling $O^\dagger O \to O^\dagger \langle O \rangle + O \langle O^\dagger \rangle - |\langle O \rangle^2|$ is exact, since the neglected fluctuation part is small ($\sim 1/M$) in the thermodynamic limit. The grand-canonical free energy for the infinite-range model is therefore given by [13]

$$F = \frac{M|m|^2}{J} - T \sum_{\alpha=1}^M \ln Tr \exp -\frac{1}{T}[(E_\alpha - \mu) b_\alpha^\dagger b_\alpha + m b_\alpha^\dagger + m^* b_\alpha], \tag{7}$$

where $m = \langle \sum_\alpha b_\alpha \rangle / J$, and the chemical potential is determined by the total number of particles as $N = -dF/d\mu$.

The hard-core repulsion between bosons restricts the Hilbert space for the single site $\alpha$ to only two states, and the above trace can be easily evaluated. Up to constant terms, the disorder-averaged free energy per site is

$$\frac{\overline{F}}{M} = \frac{|m|^2}{J} - \frac{\mu}{2} - T \int P(E) dE \ln \cosh \frac{\sqrt{(E-\mu)^2 + 4|m|^2}}{2T}, \tag{8}$$

where $P(E)$ is the normalized distribution of random energies $E_\alpha$. We will first take it to be constant for energies in the interval $[-W/2, W/2]$ and zero otherwise; this should be adequate in the strong-disorder limit $W/t >> 1$ of the electronic Hamiltonian (1). The order parameter $m \neq 0$ and the chemical potential $\mu$ are determined from the equations:

$$1 = J \int P(E) dE \frac{\tanh \frac{\sqrt{(E-\mu)^2 + 4|m|^2}}{2T}}{\sqrt{(E-\mu)^2 + 4|m|^2}}, \tag{9}$$

and

$$\frac{N}{M} = \frac{1}{2} - \frac{1}{2} \int P(E) dE \frac{(E-\mu) \tanh \frac{\sqrt{(E-\mu)^2 + 4|m|^2}}{2T}}{\sqrt{(E-\mu)^2 + 4|m|^2}}. \tag{10}$$

The transition temperature is determined by demanding that $m = 0$ in the last two equations, so that

$$1 = \frac{J}{W} \int_{-W/2}^{W/2} dE \frac{\tanh\left(\frac{E-\mu}{2T_{MF}}\right)}{E-\mu} \tag{11}$$

and
$$\frac{N}{M} = \frac{1}{2} - \frac{1}{2W} \int_{-W/2}^{W/2} dE \tanh\left(\frac{E-\mu}{2T_{MF}}\right). \quad (12)$$

These two equations in general need to be solved numerically; there are two simple limits however, which allow a transparent analytic solution. First, consider the density of bosons to be $N/M \approx 1/2$. As the solution for $T_{MF}$ in this case is finite, it immediately follows from the second equation that $\mu(T_{MF}) \approx 0$, i. e. near the center of the band of random energies. More precisely, in the strong-disorder limit when $J/W << 1$, $\mu(T_{MF}) \approx W(N/M - 1/2)$ near half-filling. The Eq. (11) then implies that

$$T_{MF} \approx const. W \sqrt{\frac{N}{M}\left(1 - \frac{N}{M}\right)} \exp\left(-\frac{W}{2J}\right), \quad (13)$$

which has the standard BCS form. In the strong disorder limit under consideration $J << W$, and consequently $T_{MF}$ is much smaller than in the clean limit. As the particle density decreases from half-filling $\mu(T_{MF})$ becomes negative, and $T_{MF}$ decreases from its maximum value. At small density the simple linear relation between the chemical potential and the density of bosons breaks down, as the system crosses over to a dilute limit of almost non-interacting bosons. At exactly $N = 0$ we see from the Eq. (12) that $T_{MF} = 0$ and $\mu(T_{MF}) \leq -W/2$. It follows from Eq. (11) that as $N \to 0$

$$\mu(T_{MF}) \to -\frac{W}{2} \coth\left(\frac{W}{2J}\right) \quad (14)$$

from above. The critical temperature when $N \to 0$ may be obtained approximately:

$$T_{MF} \approx \frac{W \exp(-W/J)}{-\ln(N/M)}. \quad (15)$$

it logarithmically slowly goes to zero with decreasing density.

In the dilute limit $N/M \to 0$, the above results for the transition temperature and the chemical potential become identical to those at the BEC of non-interacting bosons in the same random potential. If one would neglect the hard-core constraint on the occupation number of bosons, the Hamiltonian (6) would describe non-interacting particles, and its disorder-averaged spectrum could be calculated. The average density of eigenvalues in the thermodynamic limit is [14]

$$\rho(\lambda) = P(\lambda) + \frac{1}{M}\delta(\lambda + \lambda_0), \quad (16)$$

where $P(\lambda)$ is exactly the same distribution of random energies as in Eq. (8), and $-\lambda_0 < -W/2$ is the energy of the single extended bound state given by [14]

$$1 = J \int P(\lambda) \frac{d\lambda}{\lambda + \lambda_0}. \quad (17)$$

This is nothing but the Eq. (11) with $\mu$ replaced by $-\lambda_0$, in the limit when $T_{MF} \to 0$. Thus, $-\lambda_0 = \mu(T_{MF})$, as in Eq. (14). The temperature of BEC into this bound state for the non-interacting bosons would be given by

$$\frac{N}{M} = \int P(\lambda) \frac{d\lambda}{\exp(\frac{\lambda+\lambda_0}{T_{BEC}}) - 1}, \tag{18}$$

which differs from the Eq. (12) only in sign in front of unity in the denominator. This becomes negligible in the limit of small number of bosons when $T_{BEC}$ is small, and as a result, the transition temperature calculated from the Eq. (11) approaches the temperature of BEC for the non-interacting system.

The reason for finiteness of $T_{MF}$ even in the dilute limit in the infinite-range model lies in the fact that the energy of the bound state is always below the rest of the band, at least in the case of uniform distribution of random energies studied above. Pazmandi et al. [14] have also shown that in this case the bound state is spatially extended, and is given by a linear combination of macroscopically large number of localized states $|\alpha>$. This is the reason why it is able to accommodate a macroscopic number of hard-core bosons. This breaks down for distributions that tend to zero at the lower edge; for example, for the triangular distribution $P(E) = 2(E + W/2)/W^2$ for $-W/2 < E < 0$, and $P(E) = 2(W/2 - E)/W^2$ for $0 < E < W/2$, it follows from Eq. (11) that as $T_{MF} \to 0$, $\mu(T_{MF}) < -W/2$ only for $W < (2\ln 2)J$ (see Figure 2). For disorder stronger that this critical value chemical potential at the transition temperature lies in the band of localized state, and there is no condensation into any particular state. This reflects the localized nature of the states for this distribution, none of which, unlike the bound state in the weak-disorder limit, can accommodate a macroscopic number of hard-core bosons. We see that the case of uniform distribution (or more precisely, of distribution which remains finite at the lower edge) is indeed quite special; in general, the crossover to BEC in the dilute limit happens only at weak disorder. At stronger disorder the transition at $T_{MF}$ does not correspond to BEC into any particular state, but to a macroscopic number of localized states forming a condensate with a well defined phase. This is precisely the state Ma and Lee described as a "localized superconductor" [4].

## 4 Nearest-neighbor hopping

I now turn to a more realistic case of nearest-neighbor hoping of bosons and assume a fixed chemical potential at $\mu = 0$, i. e. half-filled band. The mean-field free-energy in a fixed realization of disorder is now

$$F_{MF} = \sum_{\alpha,\beta} m_\alpha^* J_{\alpha,\beta}^{-1} m_\beta - T \sum_\alpha \ln \cosh \frac{\sqrt{E_\alpha^2 + 4|m_\alpha|^2}}{2T}, \tag{19}$$

where the site-dependent order-parameter $m_\alpha$ is determined as

$$m_\alpha = J \sum_{\beta}' \frac{m_\beta}{\sqrt{E_\beta^2 + 4|m_\beta|^2}} \tanh \frac{\sqrt{E_\beta^2 + 4|m_\beta|^2}}{2T}. \tag{20}$$

The Hartree term is again neglected, but this time purely for convenience; its presence would merely make the analysis more complicated without affecting the main points. The sum in the last equation is performed only over the nearest neighbors of the site $\alpha$.

One faces a set of $M$ coupled non-linear equations, and at first it is not obvious that a non-trivial ($m_\alpha \neq 0$) superconducting solution exists at all. One can notice from the Eq. 20 however, that if it does, *all* $m_\alpha$ can be chosen to be real. It is useful to consider a toy two-site problem. At $T = 0$ the solution of the gap equations is

$$m_{1(2)} = J \sqrt{\frac{1 - (E_1 E_2/J^2)^2}{1 + (E_{2(1)}/J)^2}}. \tag{21}$$

The real solution exists only if $E_1 E_2/J^2 < 1$, i. e. if the random energy of at least one of the sites is sufficiently close to the chemical potential. In a lattice there obviously will be many nearest neighbors where this is not satisfied; nevertheless, the superconducting solution always exists due to the lattice sites with energies close to $\mu$, which then spread the long-range order throughout the whole lattice [4]. To see how this happens consider a lattice where all sites have energies in vicinity of some large energy $E = \Omega >> J$, so that the only possible solution of the gap equations is the trivial ($m_\alpha = 0$) solution. Then change the energy of a single site at $\alpha = 0$ into $E_0 = 0$. The solution at $T = 0$ in the strong disorder limit to the lowest order in $J/\Omega$ is:

$$m_0 = \frac{zJ^2}{\Omega} + O(\frac{J^3}{\Omega^2}), \tag{22}$$

at the site $\alpha = 0$, and

$$m_\alpha = J(\frac{J}{\Omega})^{n-1} + O(J(\frac{J}{\Omega})^n) \tag{23}$$

at all other sites, where integer $n$ measures the shortest lattice distance of the site $\alpha$ from the site 0 ($n = 1$ is the nearest neighbor). $z$ is the lattice coordination number. Thus a single site with the energy at the chemical potential "polarizes" immediately the whole lattice at $T = 0$, and the gap opens everywhere. The same mechanism is responsible for the existence of the non-trivial solution in a truly disordered lattice with a distribution of energies. In the thermodynamic limit, for any distribution that is finite arbitrarily close to the Fermi level (i. e., does not have a true gap there) there always will be sites with energy arbitrarily close to $\mu$. At large disorder, the solution of the gap equation will be qualitatively the same as in the above model of a two-component alloy, where the majority of sites has a large energy $\sim W$, and there are rare "seeds" with zero energy in between. Allowing the values of $J$ to vary in space can not affect the conclusion either; since hoping integrals are all necessarily positive, even if they are allowed to be random does not change the fact that the non-trivial solution of the gap-equation is positive everywhere. The ground state will therefore always be a superfluid within the mean-field theory.

Having this in mind it becomes a simple matter to obtain the critical temperature at which the non-trivial solution first becomes possible. Consider such a "seed" of long-range order at $\alpha = 0$ and its nearest neighbors; their energies will typically be of the order of $W/2$. Since $T_{MF} \sim J$, in the limit $J/W << 1$ the equations become:

$$m_0 \approx J \sum_\beta \frac{m_\beta}{|E_\beta|}, \tag{24}$$

where I assumed that both $T_{MF}$ and $m_\beta << |E_\beta|$, and

$$m_\beta \approx J \tanh \frac{m_0}{2T}. \tag{25}$$

Thus the gap at the site $\alpha = 0$ satisfies the self-consistent equation

$$m_0 \approx J^2 \sum_\beta \frac{1}{|E_\beta|} \tanh \frac{m_0}{2T}, \tag{26}$$

which allows a non-trivial solution below the critical temperature

$$T_{MF} \approx J^2 \sum_\beta \frac{1}{2|E_\beta|}. \tag{27}$$

In a given realization of the random potential $T_{MF}$ will be determined by the "seed" for which the above expression is maximal. Typical value of the mean-field critical temperature is thus

$$T_{MF} \approx \frac{zJ^2}{W}, \tag{28}$$

roughly equal to the value of the gap at the site $\alpha = 0$ at $T = 0$, and quite unlike the BCS-like expression (13) in the infinite-range model. Most importantly, $T_{MF}$ is again always finite, and vanishes only in the limit of infinite disorder. This is in accordance with the result that the ground state within the mean-field theory is always superconducting [4].

## 5  Discussion

We argued that in the negative-U strongly disordered Hubbard model the effective theory at energies much below $U$ is completely bosonic, and then proceeded to obtain the mean-field transition temperature at which the bosonic electron pairs would enter the superfluid state. True $T_c$ is always below our mean-field result, and in particular, the quantum fluctuations of the phase neglected in the mean-field theory should drive it to zero at a finite disorder. The obtained $T_{MF}$ should be understood as a crossover temperature where the superfluid order parameter begins to develop an appreciable amplitude. Our main conclusion is that the $T_{MF}$ is always finite, and vanishes only in the limit of infinite disorder.

In the pure system and at weak coupling, $T_{MF}$, or the zero-temperature superconducting gap, also measures the binding energy of an electron pair, i. e. the gap for single-electron excitations. In the strongly disordered system studied here, localization of the single particle wave-functions makes the pairs tightly bound, and it is $U >> T_{MF}$ that determines their binding energy. In the mean-field ground state at $T = 0$, the superconducting gap and the minimum energy of single-fermion excitations, while close at weak disorder, become completely different at stronger disorder (see Figure 3). Initially both are suppressed at weak disorder essentially because of the decrease in density of states at the Fermi level; however, while the spatial average of the superconducting gap continues to decrease with increasing disorder, single-fermion gap after initial suppression must begin to increase, since at very strong disorder it approaches $U$. The turning point marks the entrance into the strong-disorder limit. This divorce of the two energy scales due to disorder is clearly seen in numerical calculation of Ghosal et al. [15], where the spectral gap of Bogoliubov - de Gennes excitations (which correspond to breaking of pairs) starts increasing at large disorder. This is quite similar to what happens in the clean model with increase of attraction between fermions [11]. In sum, there are three characteristic energy scales in the disordered problem: the pair binding energy $\sim U$, the zero-temperature superconducting gap $\sim T_{MF}$, and the true $T_c$ where the phase coherence sets in, and which lies beyond the scope of the mean-field considerations.

In the simplified effective theory of the hard-core disordered bosons with infinite range hopping and with a bound distribution of random energies that vanishes at the bottom, there is a crossover to the BEC in the dilute limit if disorder is weak enough. At stronger disorder one is in the BCS "localized superconductor" limit [4], where a large number of localized states forms a condensate with a well defined macroscopic phase. In both cases the mean-field transition temperature remains finite at any disorder. There is again a close analogy between these results and the BCS-BEC crossover in the clean system of attracting fermions when the strength of attraction is varied [11]. In that case the role of hopping $J$ is played by the attractive interaction, and that of disorder $W$ by the band width $t$: in three dimensions density of states at the bottom of the band vanishes, and the bound state exist only if attractive interaction is large enough. As one increases the attraction at a fixed density of particles the system crosses from the BCS to the BEC limit, where the bound state below the band becomes macroscopically occupied. The case of uniform distribution of random energies in our problem would in this sense be similar to the case of two dimensions, where the bound state exists already for an infinitesimal attractive interaction.

It is interesting that although at first sight the nature of the BEC and that of the localized superconductor is quite different, the crossover between them in the infinite range model is completely smooth, without a phase transition. It has been argued that in the strong disorder limit, at $T = 0$, if one enters the superfluid phase by varying chemical potential, when $\mu$ is precisely at the lower edge of the band system is in the Bose-glass phase [14]. According to this scenario one could have a direct Mott insulator - superfluid (of BEC type) transition at weak disorder, and Mott insulator - Bose glass - superfluid (localized superconductor) sequence of transitions at strong disorder, depending whether there is, or there is not, the extended bound state in the spectrum. From our point of view the alleged Bose glass phase corresponds to the zero density of bosons, or more pictorially, to just a few bosons localized in the random potential. It is therefore not a macroscopic phase. At any finite density the mean-field theory yields a superfluid ground state of either BEC or localized superconductor type, with a smooth crossover between them.

The mean-field transition temperature in the model with only nearest-neighbor hopping is also always finite, and in the strong-disorder limit may be obtained by considering the opening of the superconducting gap at a particular zero-energy site. This immediately triggers the long-range order in the entire lattice, and the gap opens everywhere below $T_{MF}$. In the strong disorder limit superconducting gap vanishes exponentially fast away from these special zero-energy sites. As disorder is increased, both $T_{MF}$ and the average value of the gap decrease, but always remain finite. The low energy excitations are located at the zero-energy sites, where the gap is $\sim J^2/W$ at $T = 0$, while at farther sites the value of the gap is very close to zero. This has been observed in the numerical solution of Bogoliubov-de Gennes equations by Ghosal et al. [15], who find spatial correlations between the "superconducting islands" where the gap is substantial and the sites with small random energies. At large disorder the system becomes very much like a granular superconductor. This feature can be readily understood in the strong-disorder limit.

Evidently, the description of the superfluid-insulator quantum phase transition in the simple attractive model studied here requires an approach beyond the mean-field theory [4], [6], [7]. The mean-field theory does indicate however, that the relevant degrees of freedom for the long wave-length physics near such a transition are quantum phase fluctuations, which become enhanced at strong disorder due to a small typical amplitude of the order parameter. In ref. 7 it was shown that a disordered Bose system will turn into an insulator at $T = 0$ if the average phase stiffness at a microscopic scale is small enough. Small randomness in the stiffness, however, turns out to be irrelevant at the superfluid-Bose glass transition, at least within the $\epsilon = d - 1$ expansion [7]. Thus the model of interacting bosons in random potential should be the appropriate starting point for addressing the universal properties of the superconductor-insulator transition, at least in the strong disorder limit. We may obtain a crude estimate of the shape of the superfluid-Bose glass phase boundary at strong disorder as follows: dirty-boson Hamiltonian in Eq. 5 will have an insulating Bose-glass state at some critical disorder $W = xJ$, where $x$ is a non-universal number of order of unity. Since the random eigenstates are strongly localized, away from its maxima we may assume they go to zero exponentially, over a characteristic length (in units of lattice spacing) $\propto t/W$. The overlap integral then $J \propto \exp-(W/t)^2$, so that at strong disorder the superfluid-Bose glass phase boundary is at:

$$\frac{U}{t} \propto \exp c(\frac{W}{t})^2 \qquad (29)$$

where $c$ is yet another non-universal constant. In principle there could be a prefactor proportional to some power of $W/t$, but the main point is that at strong disorder one expects the transition to occur at a very large $U$. Thus the phase boundary in the $U-W$ plane when observed over a limited range of large $U$ should resemble a constant, in qualitative agreement with the recent Monte Carlo calculation [17].

The crucial question is if the purely bosonic nature of the superconductor-insulator transition persists at arbitrarily small $U$. It has been claimed recently [16] that at small attraction the exit from the superconducting state coincides with the appearance of the zero-energy single-fermion excitations, i. e. with breaking of electron pairs. I suspect that this feature may be a finite-$T$ artifact of the numerical simulations, in which temperature enters as a finite length of the system in the imaginary time direction. In fact, when $t = 0$ in the Hamiltonian (1) the single-particle density of states can be calculated exactly [16], and while at $T = 0$ there is always a gap equal to $U$, if $U < W/2$ density of states at zero energy becomes finite at any finite temperature. The finite temperature thus can mimic the closing of the single-fermion gap if the attractive interaction is weak enough, even though at zero temperature the gap is always present.

Another important question is validity of the bosonic theory if repulsion between electrons is included. Increasing disorder localizes the wave-functions and thus decreases the "size" of the electron pairs; this effectively increases the repulsion and may lead to pair-breaking. A precursor of this effect is the reduced screening of the Coulomb interaction due to disorder, which is believed to be an important source of reduction of order parameter amplitude or $T_c$ in real systems [9]. It is conceivable that if the repulsion is large enough pair breaking becomes the mechanism of the transition, so that the transition out of the superfluid is into an electronic glass. Another possibility is that the Bose-glass state will always set in due to quantum fluctuations when the average amplitude of the superconducting order parameter becomes sufficiently small, so that the loss of phase coherence always occurs before pairs dissociate. In this scenario, at strong repulsion Bose-glass would exist only in a narrow region between the superconductor and the electronic glass, where the electron pairs become finally broken. Studies of the interplay of the two effects are obviously required before our understanding of the superconductor-insulator transition may be considered complete.

# 6  Acknowledgment

This work has been supported by NSERC and the Faculty of Science research grant D-5765 at Dalhousie University. Part of it was also supported by Killam foundation while the author was at the University of British Columbia.

FIGURE CAPTIONS:

Figure 1: The schematic temperature-disorder phase diagram of the negative-$U$ disordered Hubbard model. While the $T_{MF}$ (dashed line) initially depends on disorder only weakly (the regime of Anderson theorem), for stronger disorder it becomes suppressed (Ma and Lee regime). $T_{MF}$ always remains finite at finite disorder. True critical temperature $T_c$ (full line) vanishes entirely due to quantum phase fluctuations. The thick line at $T = 0$ and $W > W_c$ represents the Bose glass insulator.

Figure 2: Example of the density of states in the infinite-range effective theory below (a) and above (b) critical disorder [14]. Below, there is a single eigenstate (weight 1/M) detached from the band, which is spatially extended, and available for Bose condensation at low densities. The rest of the band is localized. Above the critical disorder the extended state merges with the continuum and localizes. Nevertheless, the ground state of the system is always a superconductor.

Figure 3: Qualitative behavior of the average superconducting gap at $T = 0$ ($\overline{\Delta} \sim T_{MF}$) and the single-fermion gap (dashed line) with disorder. While the two are close at weak disorder (and not too strong $U$), at large disorder they differ by $U$.

*Present and permanent address.

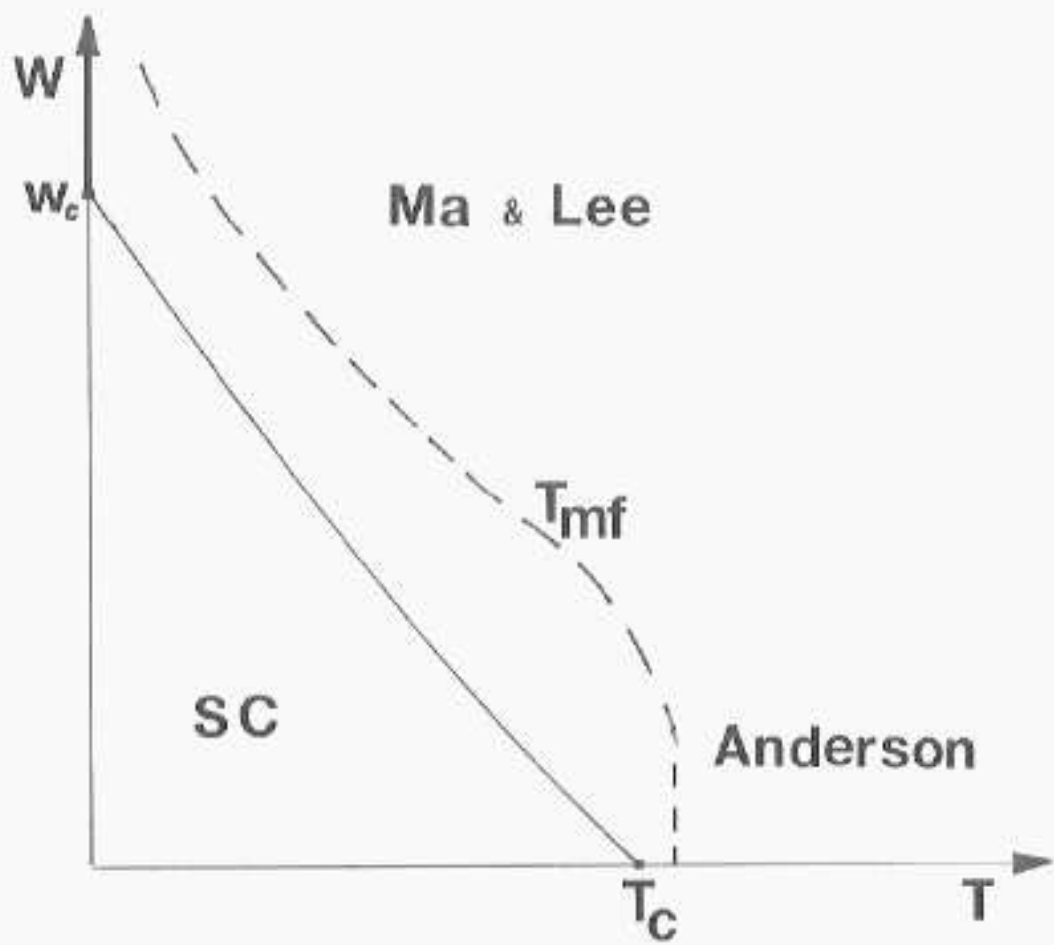

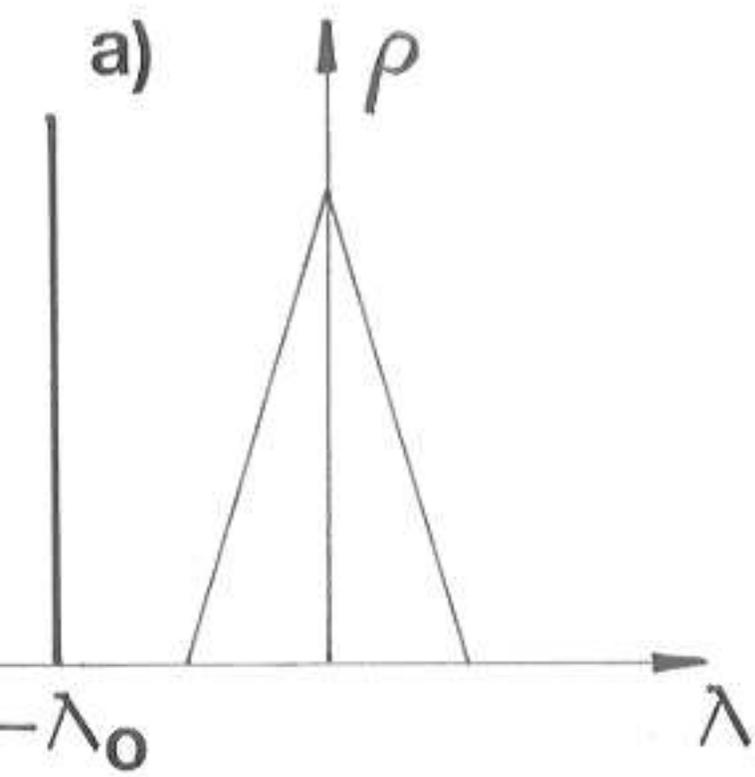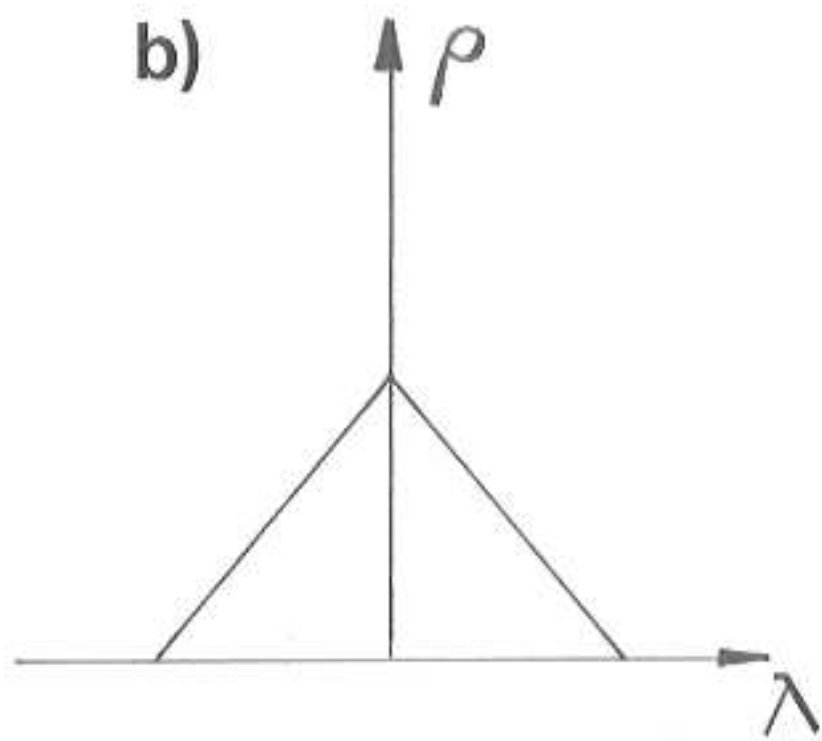

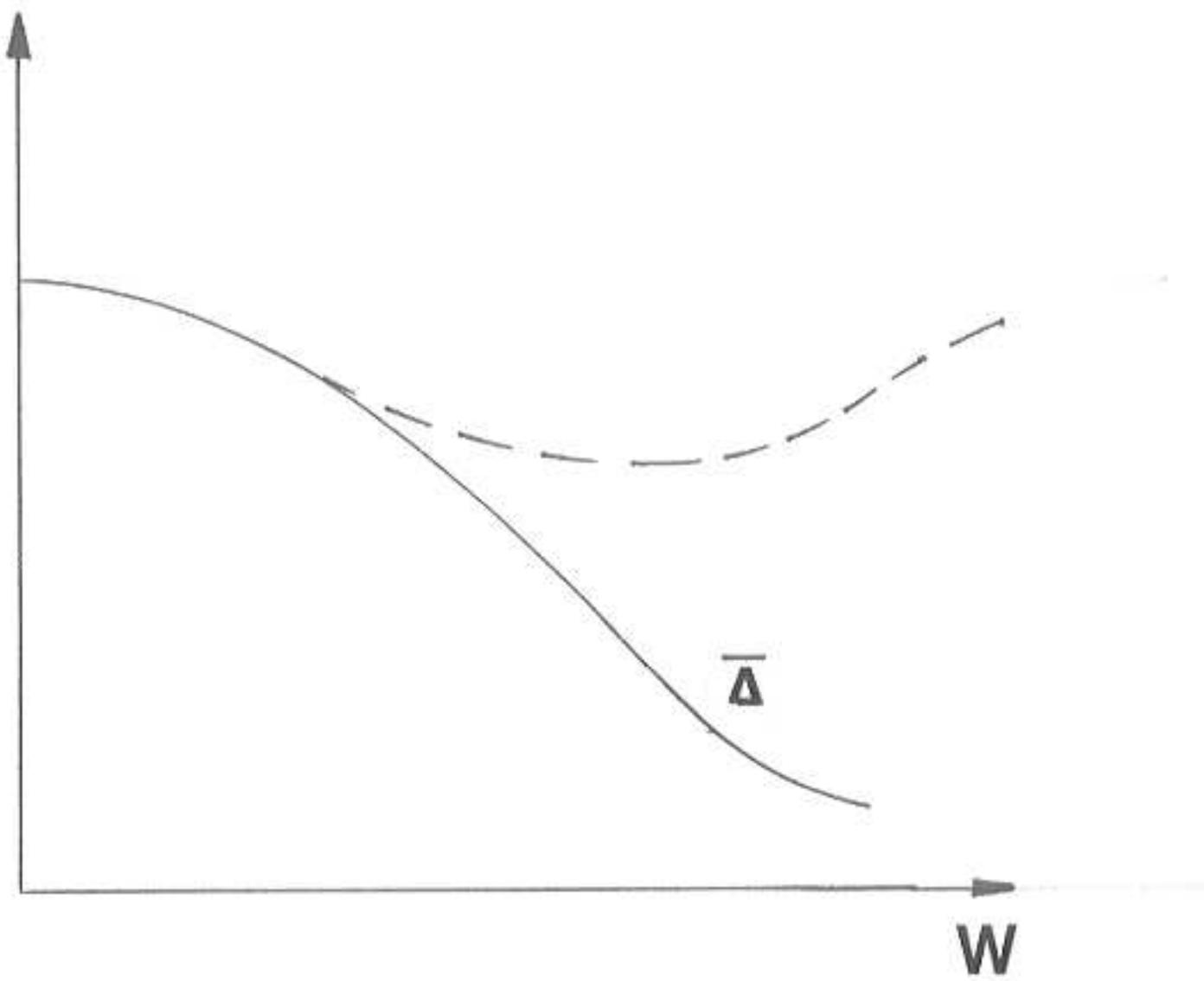
$\overline{\Delta}$

W